\documentclass[10pt]{article}

\usepackage[superscript,sort]{cite}

\usepackage{ifthen,ifpdf}
\ifpdf
	\usepackage[pdftex]{hyperref}	\hypersetup{colorlinks=true,linkcolor=black,citecolor=black,urlcolor=blue,backref=page,bookmarks=true,breaklinks=true,plainpages=false }
	\usepackage[pdftex]{graphicx}
	\usepackage[usenames,pdftex]{color}
\else
	\usepackage[colorlinks=true,breaklinks=true,plainpages=false]{hyperref}
	\usepackage{graphicx}
	\usepackage[usenames]{color}
\fi

\usepackage{amsthm,amscd,amsxtra,amsfonts,amsmath,amssymb,multirow}
\usepackage{wrapfig}
\usepackage[footnotesize]{caption}
\usepackage[tiny,compact]{titlesec}
\usepackage[textwidth=0.8in,textsize=footnotesize]{todonotes}
\usepackage{algorithm,algorithmic,extarrows}

\begin{document}

\title{Persistent homology analysis of osmolyte molecular aggregation and their hydrogen-bonding networks}

\author{
Kelin Xia$^{1,2}$ \footnote{ Address correspondences  to Kelin Xia. E-mail:xiakelin@ntu.edu.sg}, D Vijay Anand$^1$, Shikhar Saxena$^{2}$, Yuguang Mu$^{2}$ \\
$^1$Division of Mathematical Sciences, School of Physical and Mathematical Sciences, \\
Nanyang Technological University, Singapore 637371\\
$^2$School of Biological Sciences \\
Nanyang Technological University, Singapore 637371\\
}

\date{\today}
\maketitle

\begin{abstract}
Two types of osmolytes, i.e., trimethylamin N-oxide (TMAO) and urea, demonstrate dramatically different properties in a protein folding process. Even with the great progresses in revealing the potential underlying mechanism of these two osmolyte systems, many problems still remain unsolved. In this paper, we propose to use the persistent homology, a newly-invented topological method, to systematically study the osmolytes molecular aggregation and their hydrogen-bonding network from a global topological perspective. It has been found that, for the first time, TMAO and urea show two extremely different topological behaviors, i.e., extensive network and local cluster. In general, TMAO forms highly consistent large loop or circle structures in high concentrations. In contrast, urea is more tightly aggregated locally. Moreover, the resulting hydrogen-bonding networks also demonstrate distinguishable features. With the concentration increase, TMAO hydrogen-bonding networks vary greatly in their total number of loop structures and large-sized loop structures consistently increase. In contrast, urea hydrogen-bonding networks remain relatively stable with slight reduce of the total loop number. Moreover, the persistent entropy (PE) is, for the first time, used in characterization of the topological information of the aggregation and hydrogen-bonding networks. The average PE systematically increases with the concentration for both TMAO and urea, and decreases in their hydrogen-bonding networks. But their PE variances have totally different behaviors. Finally, topological features of the hydrogen-bonding networks are found to be highly consistent with those from the ion aggregation systems,  indicating that our topological invariants can characterize intrinsic features of the ``structure making" and ``structure breaking" systems.
\end{abstract}

Key words:
Persistent homology,
Molecular aggregation,
Hydrogen-bonding network,
Persistent Betti number,
Persistent entropy.
\newpage



\section{Introduction}
Among the various small molecules that nature employs to cope with the osmotic stress, urea and trimethylamine N-oxide (TMAO) are two osmolytes that attract the most attention. Urea is highly active in a variety of biological processes in the human body as well as those of other mammals and organisms. Urea belongs to a class of compounds known as chaotropic denaturants, which unravel the tertiary structure of proteins by destabilizing internal, non-covalent bonds between atoms. One of the suggested mechanism of urea-induced protein denaturation process is through an indirect effect in which urea perturbs the water-network structure \cite{rossky2008protein}. The local properties of water within solvation shell of urea molecule, have been investigated extensively, using both experimental \cite{rezus2006effect,panuszko2009effects} and molecular dynamics methods \cite{idrissi2010effect,bandyopadhyay2014molecular}. Dramatically different from urea, TMAO helps to not only stablize protein structure, but also fold intrinsically-disorder regions of proteins\cite{baskakov1999trimethylamine,baskakov1998forcing,uversky2001trimethylamine}. Even though TMAO has been employed to counteract deleterious effect of urea\cite{tseng1998natural}, its stabilization mechanism still remains elusive. It has been proposed that the oxygen atom of TMAO can form strong H-bonds with urea, thus reducing and suppressing H-bonds between urea and protein\cite{rosgen2012volume,paul2007structure}. However,  X-ray and neutron scattering experimental results show that urea-TMAO and urea-protein binding affinities are in the same order\cite{meersman2011x}. Further, molecular dynamics (MD) is used to simulate the interactions of amino acids in both urea only and urea with TMAO solutions. The results suggest that TMAO may stabilize proteins indirectly by removing urea from the protein surface\cite{ganguly2015mutual}. Even with all these models and experimental results for urea and TMAO, the detailed mechanism for their molecular aggregations and corresponding hydrogen-bonding network structures remain largely elusive.

An extensive comparison between available experimental observables and computational simulations\cite{mason:2004structure,mason:2005nanometer,Xenides:2006hydrogen,kumar:2007hydrogen} is still the most promise approach to understand the  osmolyte systems. In chemistry and biochemistry, researchers always rely on graph or network based models\cite{radhakrishnan:1991graph, dos:2004topology,Oleinikova:2005formation}, especially the spectral graph models and combinatorial graph models, to characterize the biomolecular structures, interaction networks, hydrogen-bonding network, etc\cite{Bako:2008water,da:2011hydrogen,Bako:2013hydrogen,choi2018graph,Choi:2014ionII,choi:2015ion,choi2016ion}. Graph-based descriptors including, node degree, shortest path, clique, cluster coefficient, closeness, centrality, betweenness, Cheeger constant, modularity, graph Laplacian, graph spectral, Erd\H{o}s number, percolation information, etc, have been widely used to characterize the biomolecular systems. To incorporate more geometric information into consideration, the Voronoi diagram has been proposed in the quantitative characterization of biomolecular structure, surface, volume, cavity, void, tunnels, interface, etc\cite{petvrek2007mole,edelsbrunner2005geometry,chalikian1998thermodynamic,cazals2006revisiting}. Recently, it has been used in the study of osmophobic effect\cite{smolin2017tmao}. Other than the graph and geometry models, a new model known as persistent homology has demonstrated great promising in data analysis. Different from traditional topological models, persistent homology is able to retain certain geometric information of the structure, thus it works as a bridge between geometry and topology. Persistent homology is used to study the ``shape" of data by the characterization of their intrinsic structure properties. It has been used in a variety of fields, including shape recognition \cite{DiFabio:2011}, network structure \cite{Silva:2005,LeeH:2012,Horak:2009}, image analysis \cite{Carlsson:2008,Pachauri:2011,Singh:2008,Bendich:2010,Frosini:2013}, data analysis \cite{Carlsson:2009,Niyogi:2011,BeiWang:2011,Rieck:2012,XuLiu:2012}, chaotic dynamics verification \cite{Mischaikow:1999}, computer vision \cite{Singh:2008}, computational biology \cite{Kasson:2007,YaoY:2009, Gameiro:2013}, amorphous material structures\cite{hiraoka:2016hierarchical,saadatfar:2017pore}, etc. To facilitate its application, researchers have developed many softwares, including JavaPlex \cite{javaPlex}, Perseus  \cite{Perseus}, Dipha \cite{Dipha}, Dionysus \cite{Dionysus}, jHoles \cite{Binchi:2014jholes}, GUDHI\cite{gudhi:FilteredComplexes}, etc\cite{fasy:2014introduction}. For a better visualization of the results, various models, including persistent diagram\cite{Mischaikow:2013}, persistent barcode\cite{Ghrist:2008barcodes}, and persistent landscape\cite{Bubenik:2007,bubenik:2015}, are proposed.

The application of persistent homology in biomolecular structure analysis is very promising. Currently, the model has already been used in biomolecular structure, flexibility, dynamics and function analysis\cite{KLXia:2014c, KLXia:2015a,BaoWang:2016a}. For extremely-large sized biomolecules, multiresolution and multidimensional persistent homology \cite{KLXia:2015c,KLXia:2015b} have been proposed. Weighted persistent homology has been proposed and used in the classification of different DNA types and the clustering of DNA trajectories\cite{meng2019weighted}. Further, persistent homology based machine learning and deep learning models\cite{pun2018persistent} have made enormous progress in protein-ligand binding affinity prediction\cite{cang:2017topologynet,cang:2017integration,nguyen:2017rigidity}, protein stability changes upon mutation\cite{cang:2017analysis,cang:2018representability} and toxicity prediction\cite{wu:2018quantitative}. With the features extracted from persistent homology, the accuracy of the learning model can be significantly improved. The state-of-art results for drug design, especially protein-ligand binding affinity prediction, can be achieved using the topology-based models\cite{nguyen2019mathematical,nguyen2018algebraic}.

Recently, we have applied persistent homology in the analysis of ion aggregations and hydrogen-bonding networks \cite{xia2018persistent}. Our model characterizes very well the two types of ion aggregation models, i.e., local clusters and extended ion networks. More interestingly, we have identified, for the first time, the two distinguishable topological features for the two hydrogen-bonding network systems \cite{xia2018persistent}. In this paper, we propose to use the persistent homology to analyze the topological information of molecular aggregations and their hydrogen-bonding networks. We carry out the molecular dynamic (MD) simulation of TMAO and urea in water solvent  with different concentrations and systematically analyze the simulation data. More specifically, we decompose the simulation results into two parts, i.e., osmolyte molecules and waters, and further coarse-grain TMAO, urea and water molecules as their nitrogen, oxygen and oxygen atoms, respectively. Persistent homology analysis is applied on these coarse-grained models. For the first time, we have revealed that TMAO and urea show two extremely different topological behaviors, i.e., extensive networks and local clusters. With the concentration increase, TMAO will form highly consistent large loop or circle structures, while urea tends to tightly aggregate locally with small circle structures. For TMAO hydrogen-bonding networks, their total number of loop structures consistently decreases, while at same time large-sized loop structures steadily increase. In contrast, only slight decrease of the total number of loop structures in urea hydrogen-bonding networks is observed and their large-sized loop structures increase marginally. Further, for the first time, we introduce the persistent entropy (PE) to describe the topological information of the molecular aggregation and hydrogen-bonding networks. We have found that the average PE increases with the concentration for both TMAO and urea, and decreases in their hydrogen-bonding networks. Moreover, the PE variance for TMAO and urea decreases while the PE variance for their hydrogen-bonding networks increases with the concentration. Lastly, a consistent pattern in our topological features is found for the hydrogen-bonding networks from the two osmolyte molecules and the two special ions aggregation systems of NaCl and KSCN \cite{xia2018persistent}, indicating that our topological representation characterizes the essential difference between ``structure making" and ``structure breaking" systems.

The paper is organized as follows. Section \ref{sec:method} is devoted for the introduction of persistent homology and topological barcodes for molecules. The main results are presented in Section \ref{sec:Results}. And the paper ends with a conclusion.

\section{Method}\label{sec:method}

Persistent homology is a newly-invented model deeply rooted in algebraic topology, computational topology and combinatorial topology. In persistent homology, algebraic tools, such as quotient group, homology, exact sequence, etc, are used to characterize topological invariants, including connected components, circles, rings, channels, cavities, voids, etc. We present a very brief introduction in this section and refer interested readers to papers\cite{xia2018persistent,KLXia:2014c, KLXia:2015a,Edelsbrunner:2002,Zomorodian:2005} for more details.

\subsection{Persistent homology}

\begin{figure}
\begin{center}
\begin{tabular}{c}
\includegraphics[width=0.8\textwidth]{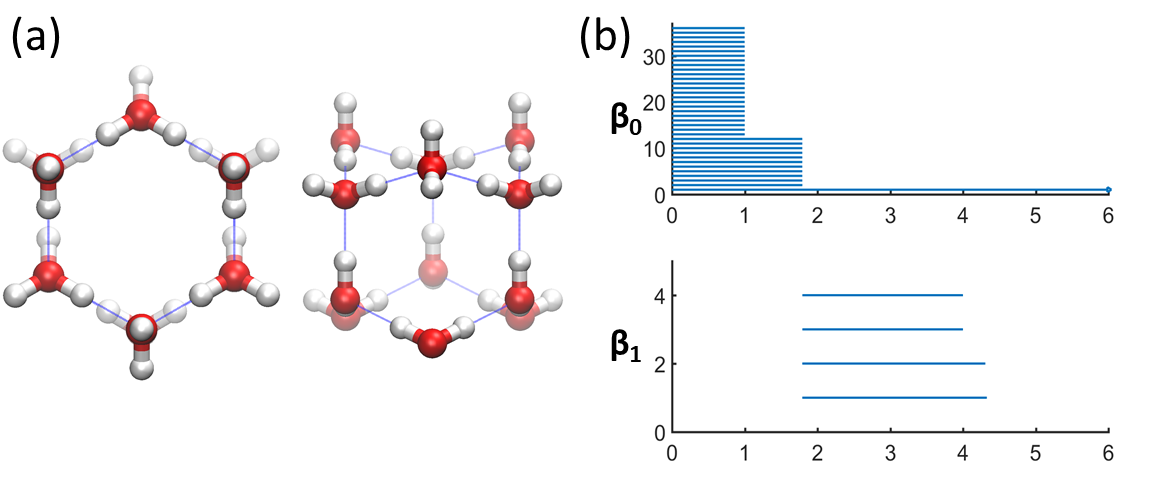}
\end{tabular}
\end{center}
\caption{ The topological representation of ice hexagonal crystal form. ({\bf a}) The ice hexagonal crystal viewed from top and side. ({\bf b}) Barcode representation for ice hexagonal crystal. The number of $\beta_0$ bars is exactly the number of atoms. There are totally 36 atoms, thus 36 $\beta_0$ bars. Usually, the length of $\beta_0$ bars represents bond length\cite{KLXia:2014persistent}. For the ice crystal structure, there are two types of $\beta_0$ bars with length 1.00 \AA~ and  1.79 \AA, which are exactly the H-O covalent and H-O non-covalent bond lengths, respectively. Further, $\beta_1$ bars are related to the cycle, loop or ring structures. There are 5 ring structures, 2 larger ones on the top/bottom and 3 smaller ones on the side, in the crystal structure. Algebraically, one small ring can be expressed as a combination of the rest 4 rings. Therefore, there are totally 4 $\beta_1$ bars, 2 longer ones from top and bottom and 2 smaller ones from the side.
}
\label{fig:Crystal_lattice}
\end{figure}

Persistent homology provides a multiscale representation of topological invariants from simplicial complexes. Generally speaking, a network or graph is a special kind of simplicial complex with only 0-simplexes (nodes or vertexs) and 1-simplexes (edges). Simplicial complexes can have higher-dimensional components\cite{giusti2016two}, such as 2-simplexes and 3-simplexes, which can be geometrically viewed as triangles and tetrahedrons, respectively. Further, network models focus on local or combinatorial measurements, including degree, shortest path, clique, cluster coefficient, closeness, centrality, betweenness, Cheeger constant, modularity, graph Laplacian, graph spectral, Erd\H{o}s number, percolation information, etc. Persistent homology studies the topological invariant called Betti numbers, which include $\beta_0$, $\beta_1$, $\beta_2$, etc. Geometrically, $\beta_0$ counts the number of connected components, $\beta_1$ is for the number circles, rings or loops, and $\beta_2$ counts the number of voids or cavities.

The key concept in persistent homology is the filtration. For point cloud data, we can associate each point with an identical-sized sphere. Filtration parameter can be defined as the size of sphere radius (or diameter). With the increase of filtration value, these spheres will enlarge and further overlap with each other to form simplexes. Roughly speaking, an edge between two points is formed when the two corresponding spheres overlap. A triangle is formed when each two spheres (of the three corresponding spheres from triangle vertices) overlap. And a tetrahedron is formed when each three spheres (of the four corresponding spheres from tetrahedron vertices) overlap. At each filtration value, all the simplexes, i.e., vertices, edges, triangles, tetrahedrons, form a simplicial complex. The topological invariants, i.e., Betti numbers, can be calculated.  Through a systematical variation of a filtration parameter, a series of simplicial complexes from different scales are generated. Some topological invariants can persistent longer in these simplicial complexes, while others quickly disappear when the filtration value changes. In this way, the ``lifespan" of the topological invariants (circles, loops, etc) in these simplicial complexes provides a natural geometric measurement. More specifically, the lifespan, known as the persistence, measures how ``large" are the circles, loops and voids in the system.

Figure \ref{fig:Crystal_lattice} illustrates the persistent homology analysis of an ice hexagonal crystal. The information of topological invariants is represented as a barcode (Figure \ref{fig:Crystal_lattice}({\bf b})). Each bar represents an individual topological invariant. The invariant $\beta_0$, representing the number of individual components, is plotted in the upper figure. Invariant $\beta_1$, denoting circle or ring structure, is shown in the lower figure. For both figures, the x-axis represents the filtration size, i.e., the diameter of the sphere with unit in Angstrom(\AA). The y-axis is the index of topological invariant. Mathematically, $\beta_0$ represents the total number of individual components in the systems. It can be seen that there are totally 36 atoms in the hexagonal crystal structure. When the filtration value goes beyond 1.0 \AA, $\beta_0$ reduces to 12, meaning these 36 atoms formed into 12 clusters. The filtration value 1.0 \AA~ is exactly the covalent bond length between the H and O atom of the same water molecule and each cluster is exactly a water molecule. When filtration value goes beyond 1.9 \AA, $\beta_0$ further reduces to 1, meaning that all atoms are very ``connected". The filtration value 1.9 \AA~ is exactly the non-covalent bond length of the H and O atom of two adjacent water molecules in the crystal. In the meantime, $\beta_1$ barcode begins to appear. Essentially, there are 5 ring structures, 2 larger ones on the top and bottom and 3 smaller ones on the side. Algebraically, one small ring can be expressed as a combination of the rest 4 rings. Therefore, there are totally 4 $\beta_1$ bars, 2 longer ones from top and bottom and 2 smaller ones from the side. These $\beta_1$ bars disappear when the filtration size is larger than around 4.0 \AA~and 4.2 \AA, when the 2-simplex, i.e., triangles, are formed and cover the whole circle. Geometrically, the length of $\beta_1$ bars is roughly the size of the circle. Since the barcode representation provides a very good description of molecular structures, it can be used as a ``fingerprint" for structure characterization, identification, and analysis \cite{KLXia:2014persistent,ZXCang:2015,KLXia:2015b,KLXia:2015a,KLXia:2015c,KLXia:2015d,xia2015multiresolution,cang:2017topologynet,cang:2017integration}.

The filtration value, at which the invariant appears or disappears, is called birth time (BT) and death time (DT), respectively. In this way, each invariant has a ``lifespan", i.e., barcode length (BL), defined by its birth and death time. Essentially, the bar length provides a geometric measurement of the topological invariant. The pairs of BTs and DTs from the persistent homology analysis can be represented as barcodes as in Figure \ref{fig:Crystal_lattice}. We can denote them as follows,
\begin{eqnarray}
\{ L_{k,j}=[a_{k,j}, b_{k,j}] | k=0,1,...; j=1,2,3,....,N_k \},
\end{eqnarray}
where parameter $k$ represents the $k$-dimension. For data points from Euclidean space, normally we only consider $k=0,1,2$. Parameter $j$ indicates the $j$-th topological invariant (barcodes) and $N_k$ is the number of $\beta_k$ topological invariant. The $a_{k,j}$ and $b_{k,j}$ are BTs and DTs. For simplification, we can define the set of $k$-th dimensional barcodes as,
$$ L_{k}= \{ L_{k,j}, j=1,2,3,....,N_k\}, \quad k=0, 1, ....$$

\paragraph{Persistent Betti functions}
Based on the persistent homology results, different functions are proposed to represent or analyze the topological information\cite{Edelsbrunner:2002,Carlsson:2009,bubenik:2015,Chintakunta:2015}. The persistent Betti number (PBN) is one of them, it is defined as the summation of all the $k$-th dimensional barcodes,
\begin{eqnarray}\label{eq:PBN}
f(x;L_{k})= \sum_{j} \chi_{[a_{k,j},b_{k,j}]}(x), \quad k=0, 1, ....
\end{eqnarray}
Function $\chi_{[a_{k,j}, b_{k,j}]}(x)$ is a step function, which equals to one in the region $[a_{k,j}, b_{k,j}]$ and zero otherwise. This equation transforms all the $k$-dimensional barcodes into a one-dimensional function. We can also define an average persistent Betti number as follows,
\begin{eqnarray}\label{eq:sPBN}
f(x;L_{k})= \frac{1}{N}\sum_{j} \chi_{[a_{k,j},b_{k,j}]}(x), \quad k=0, 1, ....
\end{eqnarray}
Here $N$ can be the total number of atoms, $\beta_k$ bar number, etc.

\paragraph{Persistent entropy}

To measure the disorder of a system, persistent entropy has been proposed\cite{Merelli:2015topological,Chintakunta:2015,Rucco:2016,Xia:2018multiscale}. For the $k$-dimensional barcodes, it is defined as,
\begin{eqnarray}\label{eq:filtrationM}
S_k=\sum_j^{N_k} - p_{k,j} ln(p_{k,j}), \quad k=0, 1, ...,
\end{eqnarray}
with the probability function,
\begin{eqnarray}\label{eq:pi}
p_{k,j}=\frac{b_{k,j}-a_{k,j}}{\sum_j (b_{k,j}-a_{k,j})}, \quad k=0, 1, ....
\end{eqnarray}
The expression of PE can be simplified as follows,
\begin{eqnarray}\label{eq:MPE}\nonumber
S_k=ln\left(\sum_j^{N_k} (b_{k,j}-a_{k,j})\right)- \frac{\sum_j^{N_k}\left( (b_{k,j}-a_{k,j}) ln(b_{k,j}-a_{k,j}) \right)}{\sum_j^{N_k} (b_{k,j}-a_{k,j})}, \\ \quad k=0, 1, ....
\end{eqnarray}

It should be noticed that PE is different from the general entropy used in molecular dynamic simulation. Generally speaking, PE characterizes the ``topological orderless". For instance, for a crystal structure, it has very consistent three-dimensional structure, thus a highly regular barcode and a large PE value. Topologically, a large PE value means that the structure is very regular and ``lattice-like" with consistent connection patterns, while a lower PE value means the components in the structure are randomly distributed with no consistent patterns.

\subsection{Topological fingerprints for two types of osmolytes}

\begin{figure}
\begin{center}
\begin{tabular}{c}
\includegraphics[width=0.6\textwidth]{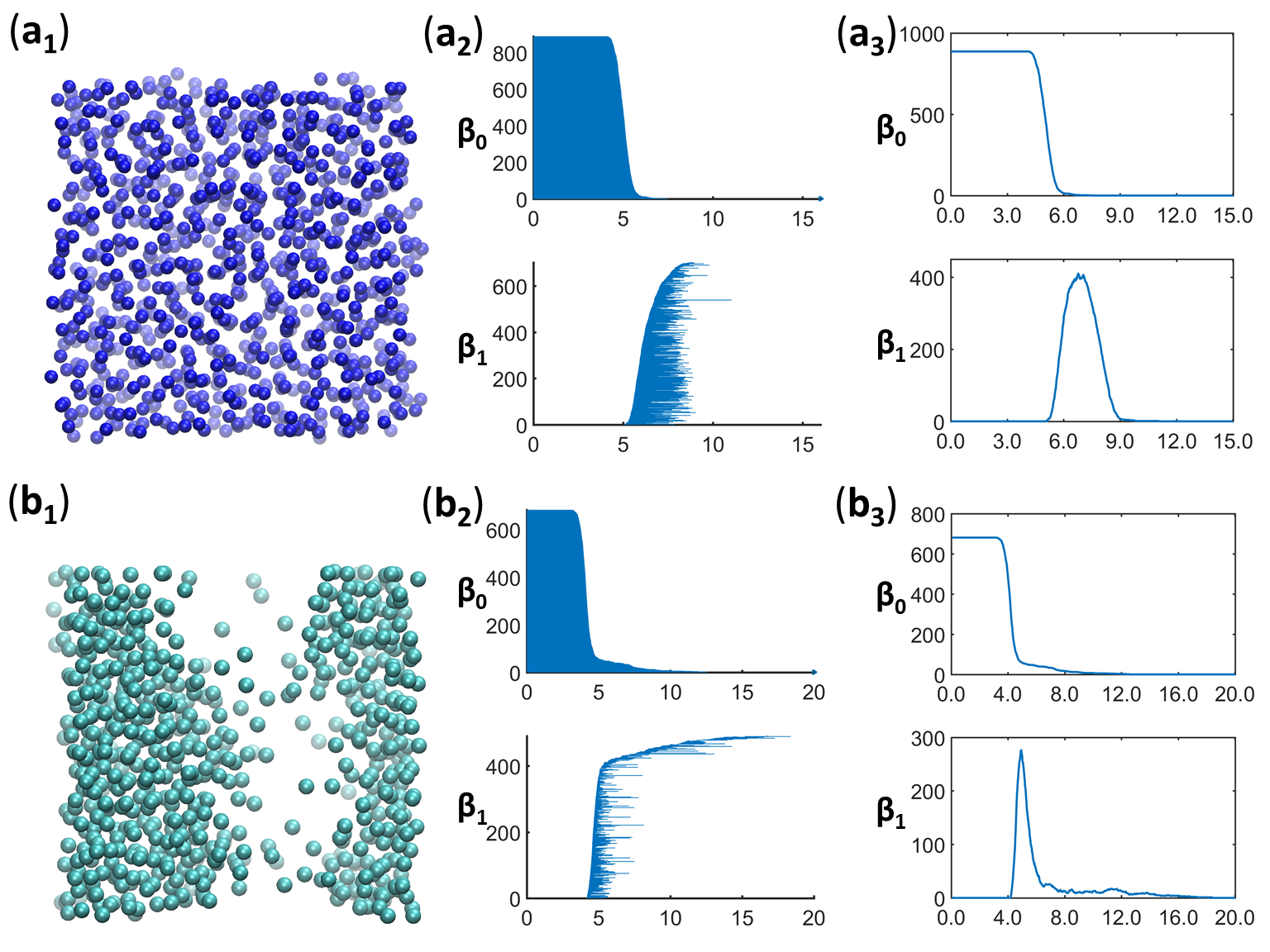}
\end{tabular}
\end{center}
\caption{ The topological fingerprints for the two types of osmolytes, i.e., TMAO ({\bf a})  and urea ({\bf b}). The two configurations from MD simulation of 8 M TMAO and urea in water solvent are considered. The TIP4P water model is considered in this case. TMAO and urea molecules are coarse-grained as their nitrogen and oxygen atoms, respectively. ({\bf a}$_1$) The aggregation of TMAO molecules. ({\bf a}$_2$) The barcode representation of TMAO aggregation topology. ({\bf a}$_3$) The PBN for TMAO aggregation. ({\bf b}$_1$) The aggregation of urea molecules. ({\bf b}$_2$) The barcode representation of urea aggregation topology. ({\bf b}$_3$) The PBN for urea aggregation. It can be seen that TMAO and urea have dramatically different way of aggregation. Roughly speaking, TMAO molecules aggregate ``loosely" with much larger local ring structures, while urea molecules tightly pack together locally.
}
\label{fig:hydrogen_mol}
\end{figure}

It is known that TMAO and urea are two types of osmolytes that have very different properties in protein folding process. The urea tends to ``destroy" the folding processes while TMAO stabilizes them. Unlike previous network or graph based models, we use persistent homology to analyze the global topology for the molecular aggregation of TMAO and urea. Stated differently, we focus on the global topology that can be viewed as the equilibrium state in which all the local interactions between molecules and molecules with waters balance. As an example, two configurations from MD simultions of TMAO and urea in 8M solution are considered in Figure \ref{fig:hydrogen_mol}. The TIP4P water model is used and the TMAO and urea molecules are coarse-grained as their nitrogen and oxygen atoms, respectively. More detailed MD setting can be found in the Section \ref{sec:Results}. The aggregation configurations, barcodes, and PBNs are demonstrated in Figure \ref{fig:hydrogen_mol}. It can be seen that TMAO and urea have dramatically different ways of aggregation. Roughly speaking, TMAO molecules aggregate ``loosely" with much larger local ring structures, while urea molecules closely and neatly pack together locally. The barcode representation characterizes the topological behavior very well. It can be seen that, in comparison with TMAO, urea has more short $\beta_1$ bars at the early stage of filtration (around 5 \AA), meaning more locally compact structures. On the other hand, urea has various $\beta_1$ bars even after 10 \AA, meaning topological structures at more global scale. In contrast, TMAO $\beta_1$ bars are highly organized and are all concentrated in the filtration range from 5 \AA~ to 10 \AA, meaning that TMAO has very regular topological network structures.

\subsection{Topological fingerprints for hydrogen-bonding networks}

\begin{figure}
\begin{center}
\begin{tabular}{c}
\includegraphics[width=0.6\textwidth]{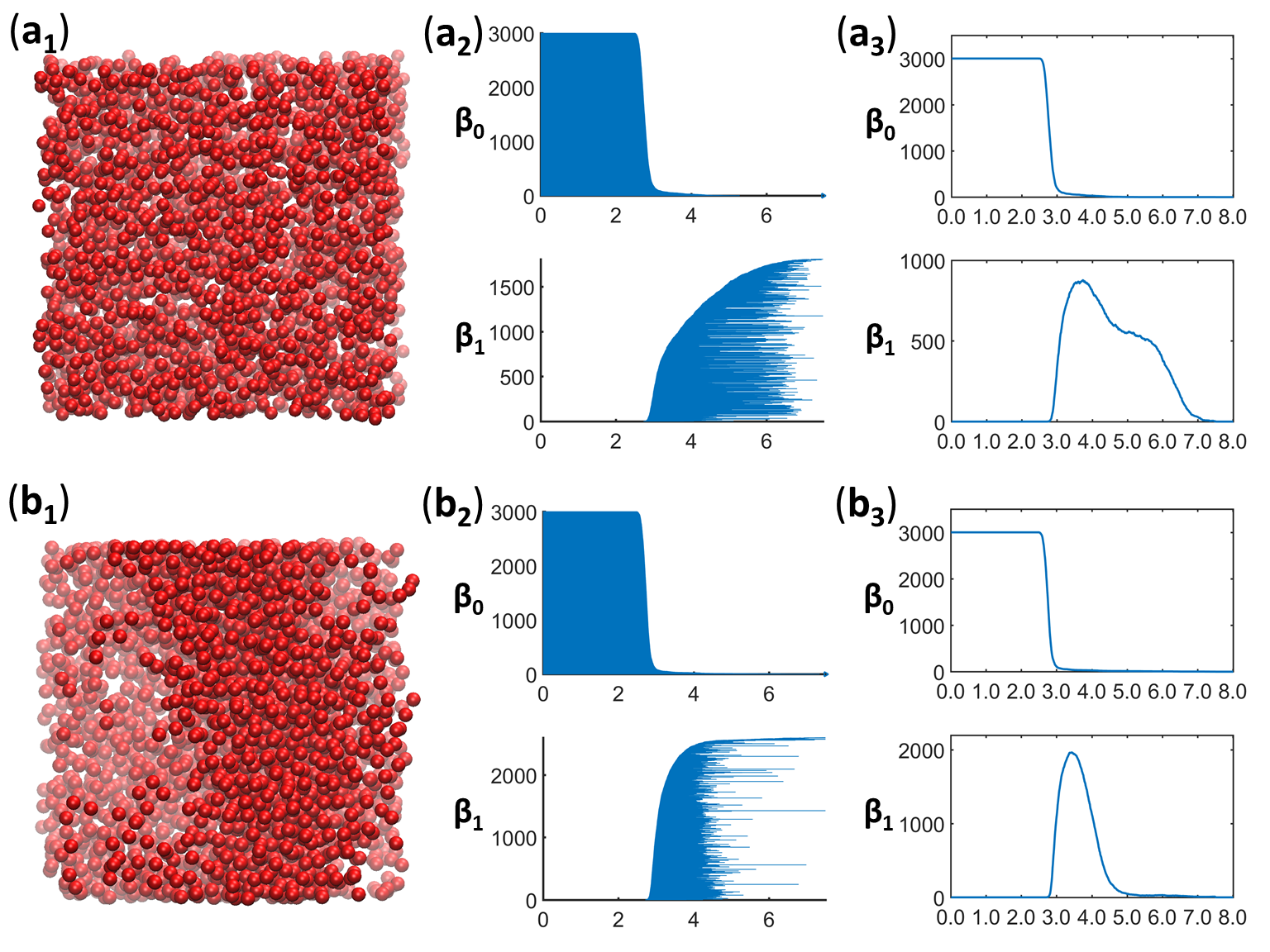}
\end{tabular}
\end{center}
\caption{ The topological fingerprint for the hydrogen-bonding networks from the two types of osmolytes, i.e., TMAO ({\bf a})  and urea ({\bf b}). The two water configurations from MD simulations (TIP4P water model) of TMAO and urea in 8M solution are considered. The water molecules are coarse-grained as their oxygen atoms. ({\bf a}$_1$) The aggregation of water molecules in TMAO system. ({\bf a}$_2$) The barcode representation of the hydrogen-bonding network from TMAO. ({\bf a}$_3$) The PBN for the hydrogen-bonding network from TMAO.({\bf b}$_1$) The aggregation of water molecules in urea systems. ({\bf b}$_2$) The barcode representation of the hydrogen-bonding network from urea. ({\bf b}$_3$) The PBN for the hydrogen-bonding network from urea.
}
\label{fig:hydrogen_network}
\end{figure}

More interestingly, we can explore the TMAO and urea hydrogen-bonding networks. The two water configurations from MD simulations (TIP4P water model) of 8 M TMAO and urea in water solvent are considered. The water molecules are coarse-grained as their oxygen atoms.  Figure \ref{fig:hydrogen_network} illustrates the aggregation configurations, barcodes, and PBNs. It can be seen that two hydrogen-bonding networks demonstrate dramatically different behaviors. Even though the two systems all have 3000 water molecules, their $\beta_1$ numbers differ a lot. Clearly, TMAO hydrogen-bonding network has less $\beta_1$ bars, meaning there are fewer circle or loop structures within the network. In the meanwhile, more longer $\beta_1$ bars are found in TMAO hydrogen-bonding network, indicating a large number of large-sized circles or loops. The very different barcode fingerprints indicate two different types of hydrogen-bonding network topology for the two hydrogen-bonding network systems.

\section{Results and discussions}\label{sec:Results}

In this section, we carry out detailed MD simulation to explore the topological structural properties for TMAO and urea molecular aggregations and the related hydrogen-bonding networks. The two molecular systems in eight different concentrations are considered. Further, we use persistent barcodes, persistent Betti number and persistent entropy to analyze their topological invariants and systematically compare their behaviors.
\begin{figure}
\begin{center}
\begin{tabular}{c}
\includegraphics[width=0.6\textwidth]{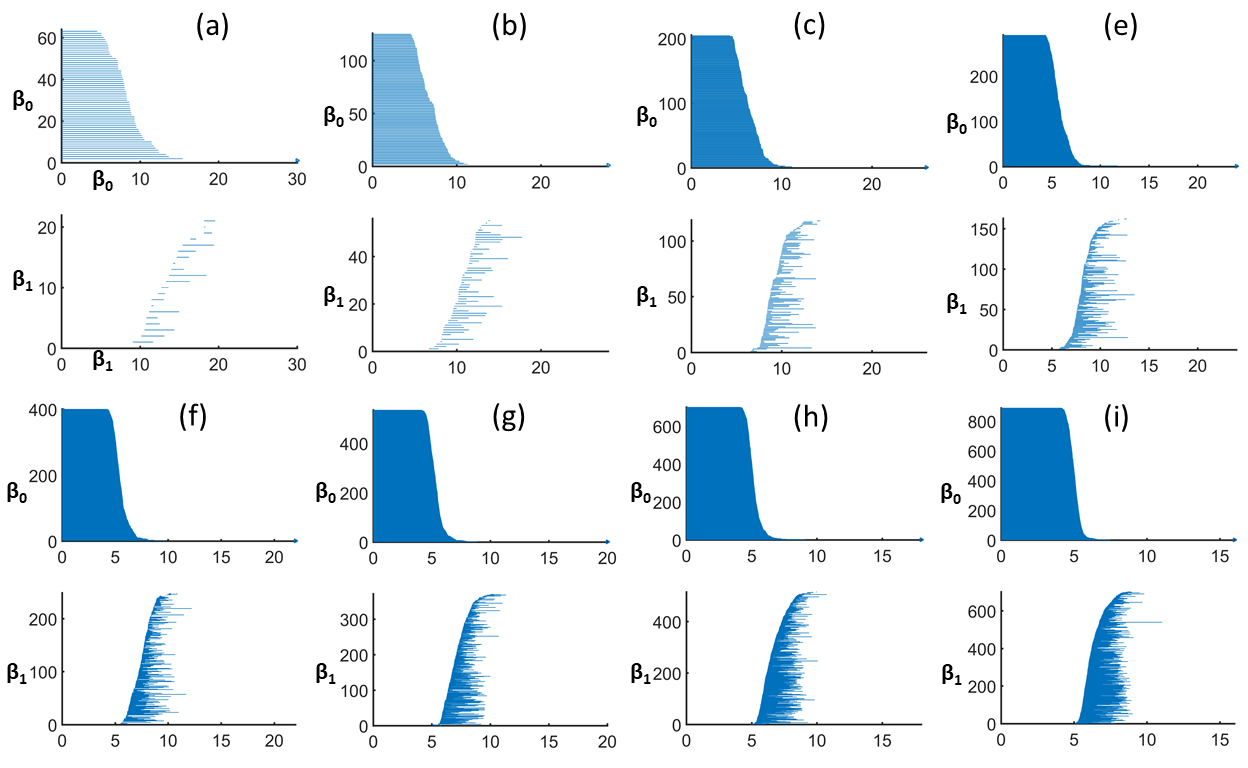}
\end{tabular}
\end{center}
\caption{The persistent barcodes for TMAO aggregation. The coarse-grained representation of TMAO as its nitrogen atom is considered. The last configuration of the MD simulation is used.  From (${\bf a}$) to (${\bf i}$) are persistent barcodes for TMAO system at ion concentration 1 M to 8 M, respectively. It can be seen that, the total bar numbers of $\beta_0$ barcodes are systematically increase from (${\bf a}$) to (${\bf i}$), indicating the increase of TMAO atoms. The $\beta_1$ barcodes are also systematically increased, meaning more and more ring/loop structures are formed. The range of $\beta_1$ barcodes shifts from the domain of [5, 20] \AA, to a much confined region of [5, 12] \AA~ with the concentration increase.
}
\label{fig:tmao_4p_ions}
\end{figure}

\subsection{Data generation and MD simulation}
The molecular dynamic simulations are performed using GROMACS-5.1.2 \cite{abraham2015gromacs,berendsen1995gromacs}. For TMAO, the Kast model \cite{kast2003binary} is adopted whereas the urea model from AMBER package\cite{pearlman1995amber} is considered along with the three point (TIP3P) and four point (TIP4P-EW) \cite{horn2004development} water model. Pure water and 1M to 8M concentrations of urea and TMAO are simulated in both water models for 100 ns each. The number of water molecules are kept to be 3000 in all the cases. Initial configurations are built by randomly placing the urea/TMAO molecules using insert-molecules utility in GROMACS followed by random insertion of 3000 water molecules in a cubic simulation box. Thus, each simulation box is comprised of 3000 molecules of water along with variable number of urea and TMAO molecules according to their concentration. Equilibration is performed first under NVT conditions (Temperature = 300 K) for 10 ps and then under NPT conditions for 100 ps using 2 fs time step, Berendsen thermostat $\tau$ = 0.1ps) and barostat ($\tau$= 2 ps). The bonds and the angles are constrained by LINCS algorithm \cite{hess1997lincs}. Calibration between the box volume and the number of solute (urea/TMAO) molecules needed to achieve the required concentration is performed using short simulation of 100 ps in NPT ensemble without any constraints. Three repeats of production run are then carried out under NPT conditions for 100 ns with Berendsen thermostat (Temperature = 300 K, $\tau$ = 0.1 ps)  Parrinello-Rahman barostat (Pressure = 1 bar, $\tau$ = 2 ps) and using a time step of 2 fs. A leap-frog algorithm is used to integrate Newton's equation of motion. The cut-off for both van der Waals (VDW) interaction and short-range electrostatic interaction are set to 1.0 nm, and a Particle mesh Ewald (PME) \cite{essmann1995smooth} method is employed to deal with the long-range electrostatic interactions. Configurations are output every 1 ps. 
For each simulation, we take 101 frames equally from the simulation trajectory. Our persistent homology analysis is based on these 101 frames.

\subsection{Topological modeling for two types of osmolytes}

To have a general idea of the topological variation of the TMAO aggregation with the increase of ion concentration, we take the last frame of each simulation and perform the persistent homology analysis. Again, TMAO is coarse-grained by using only its nitrogen atom. Figure \ref{fig:tmao_4p_ions} illustrates the barcodes for TMAO aggregations in eight different concentrations. Figure \ref{fig:tmao_4p_ions} (${\bf a}$) to (${\bf i}$) are persistent barcodes for TMAO system at ion concentration 1 M to 8 M, respectively. It can be seen that, the total bar numbers of $\beta_0$ barcodes are systematically increase from Figure \ref{fig:tmao_4p_ions} (${\bf a}$) to (${\bf i}$), indicating the increasing numbers of TMAO atoms. The $\beta_1$ barcodes are also systematically increased, meaning more and more ring/loop structures are formed. More importantly, the range of $\beta_1$ barcodes shifts from the domain of [5, 20] \AA, to a much confined region of [5, 12] \AA~ with the concentration increase, indicating that the size of loop structures becomes more and more uniform.

\begin{figure}
\begin{center}
\begin{tabular}{c}
\includegraphics[width=0.6\textwidth]{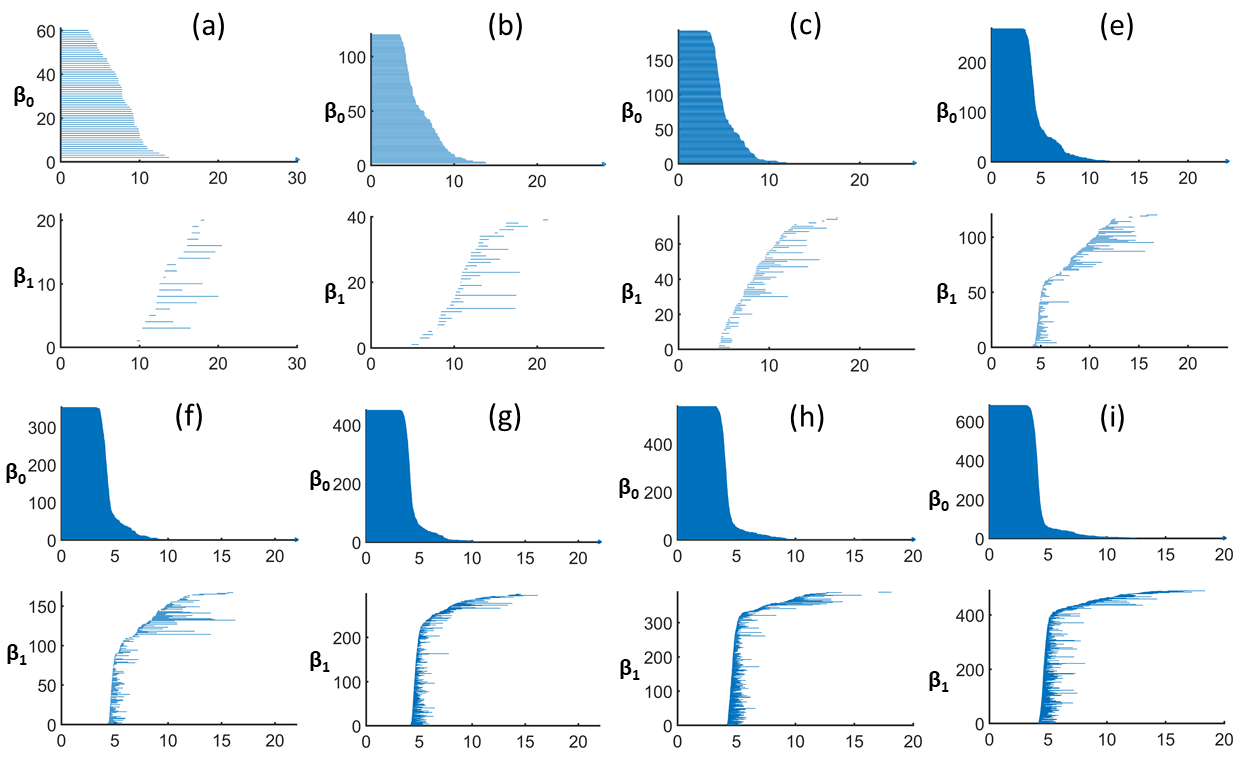}
\end{tabular}
\end{center}
\caption{The persistent barcodes for urea aggregation. The coarse-grained representation of urea as its oxygen atom is considered. The last configuration of the MD simulation is used. From (${\bf a}$) to (${\bf i}$) are persistent barcodes for urea system at ion concentration 1 M to 8 M, respectively. It can be seen that, the total bar numbers of $\beta_0$ barcodes systematically increase from (${\bf a}$) to (${\bf i}$), indicating the increasing number of urea atoms. The $\beta_1$ barcodes are also systematically increased, meaning more and more ring/loop structures are formed. Different from TMAO barcodes, as the concentration increase, most of $\beta_1$ barcodes gradually concentrate around the range of [4, 7] \AA, in the mean time, a long ``tail" of barcodes forms from 5\AA~ to around 18 \AA.
}
\label{fig:urea_4p_ions}
\end{figure}

We also check the topological fingerprints for urea aggregation. The coarse-grained representation of urea as its oxygen atom is considered. The last configuration of the MD simulation is used. Figure \ref{fig:urea_4p_ions} illustrates the barcodes for urea aggregations in eight different concentrations. Figure \ref{fig:urea_4p_ions} (${\bf a}$) to (${\bf i}$) are persistent barcodes for urea system at ion concentration 1 M to 8 M, respectively. It can be seen that, the total bar numbers of $\beta_0$ barcodes are systematically increase from Figure \ref{fig:urea_4p_ions} (${\bf a}$) to (${\bf i}$), indicating the increase of urea atoms. The $\beta_1$ barcodes also systematically increase, meaning more and more ring/loop structures are formed. Different from TMAO barcodes, with the concentration increase, most of $\beta_1$ barcodes gradually concentrate around the range of [4, 7] \AA, in the mean time, a long ``tail" of barcodes forms from 7 \AA~ to around 18 \AA.

\begin{figure}
\begin{center}
\begin{tabular}{c}
\includegraphics[width=0.8\textwidth]{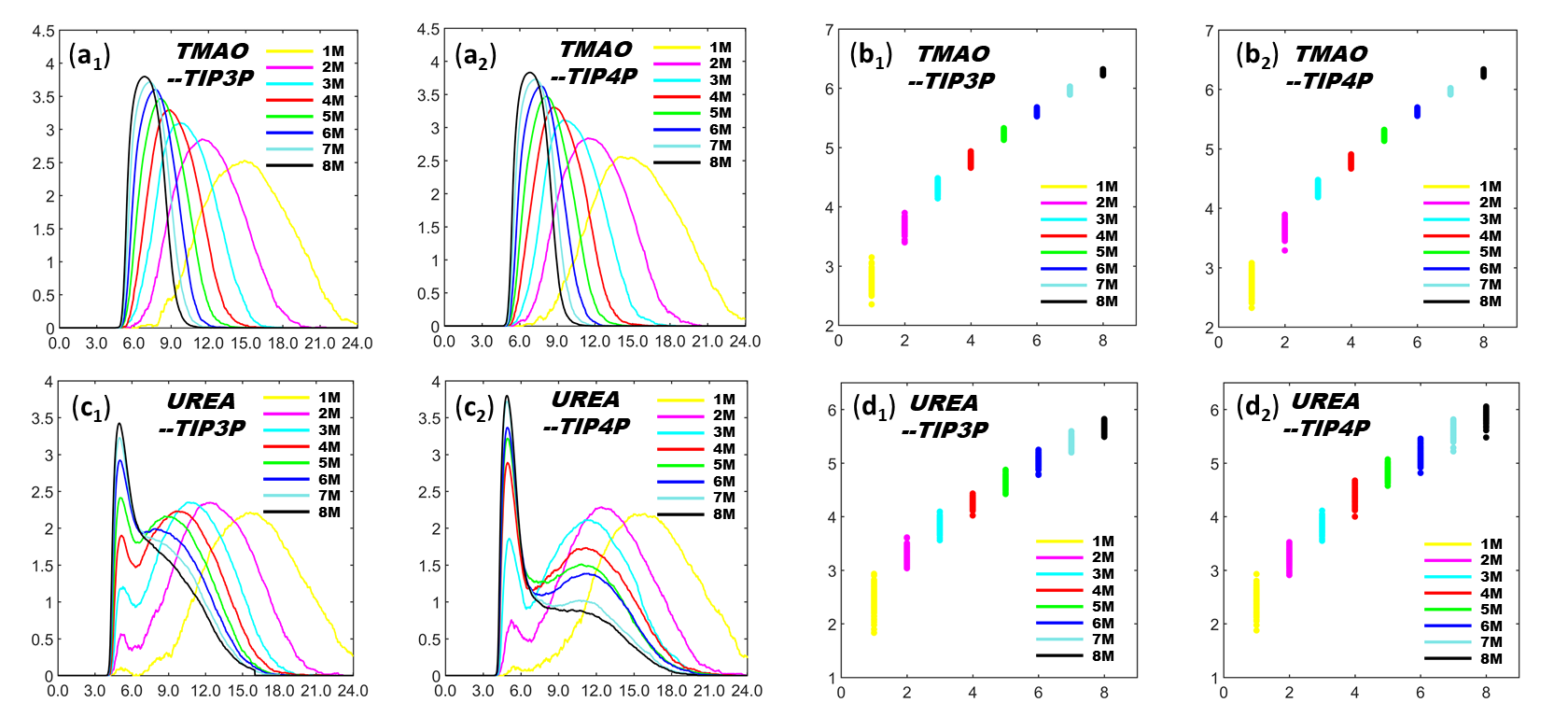}
\end{tabular}
\end{center}
\caption{ The comparison of average PBNs and persistent entropy for TMAO and urea aggregations. Two water models, i.e., TIP3P and TIP4P, are considered. The PBNs are averaged over the total number of atoms and configuration numbers. It can be seen that PBNs share very similar patterns in both TIP3P and TIP4P models. More importantly, the aggregation pattern of TMAO (${\bf a}$) is dramatically different from urea (${\bf c}$). With the concentration increase, the center of TMAO PBNs shifts from around 15\AA~ to 7\AA. In the meantime, the peak value of PBNs consistently increases. In contrast, urea PBNs gradually evolve into two peaks, one appears at around 5\AA~ and other shifts from 15 \AA~ to 8\AA~ and dies down gradually. The persistent entropies for TMAO (${\bf b}$) and urea (${\bf d}$) molecules. Totally 101 configurations for each concentration are considered. It can be seen that the PEs share very similar patterns in both TIP3P and TIP4P models. With the concentration increase, average PEs for both TMAO and urea consistently increase. However, the variance of TMAO PEs decreases much faster that urea PEs, indicating that TMAO is much more stable than urea in terms of their topological structures.
}
\label{fig:PBN_mol}
\end{figure}

All the above results are based on special configurations (last frames) from the MD simulations. To facilitate a systematical comparison of TMAO and urea molecular aggregation in the entire MD simulation, we consider their average PBNs. To be more specific, for each frame we calculate the average PBN using Eq.(\ref{eq:sPBN}) with $N$ the total number of atoms. For each ion concentration, we then summarize the average PBNs over the 101 frames and divide them by 101. Figure \ref{fig:PBN_mol} illustrates the final PBNs for both TMAO and urea in different concentrations. Firstly, we compare the PBNs from two different water models, i.e., TIP3P and TIP4P. It can be seen that the PBNs from both models share very similar patterns. More specifically, PBNs for TMAO in both models are almost identical with same peak values and general shapes. In contrast, PBNs for urea have similar shapes but their values at critical points (peaks, valleys, etc) are not exactly the same and sometimes can differ by as large as 0.3. Secondly, the aggregation behavior of TMAO is dramatically different from urea. TMAO PBNs have only one peak. With the concentration increase, the peak of TMAO PBNs shifts from around 15 \AA~ to 7 \AA. In the meantime, the peak value of PBNs consistently increases. In contrast, urea PBNs has only one peak in low concentrations, and gradually evolves into two peaks in high concentrations. Among the two peaks, one appears at around 5\AA~ and other shifts from 15 \AA~ to 8\AA~ and dies down gradually. Geometrically, the two-peak pattern indicates the existence of both local and global topological properties of urea. More specifically, urea has both local aggregation structures and also large global loop structures. In comparison, TMAO is much more regular and uniform. With the concentration increase, TMAO molecules become more and more densely packed with highly organized structures.

We calculate the persistent entropies for the two aggregation systems. Totally 101 configurations for each concentration are considered and results are demonstrated in Figure \ref{fig:PBN_mol}. Similarly, PEs share very similar patterns in both TIP3P and TIP4P models. The average PEs for both TMAO and urea consistently increase with the concentration increase. At the same time, for each concentration, various PEs are obtained and they have very different variances. Table \ref{tab:PEs} demonstrates the variances of PEs for both TMAO and urea in different concentrations. It can be seen that the variances for TMAO and urea all decrease with the concentration. However, PE variances for TMAO are decreasing much faster than urea. The ratio between the variance of concentration at 1 M and 8 M for TMAO is around 50 and 70 for TIP3P and TIP4P models, respectively. While the ratio for urea is around 10 and 4 for TIP3P and TIP4P models, respectively. In this way, even though TMAO and urea have similar variance (only 1 to 2 times difference) at concentration 1 M, they differ greatly at concentration 8 M. Variance for urea is about 14 and 21 times larger than TMAO at 8 M. Since a larger PE means more regular topological structures, the variance results mean that both TMAO and urea systems become more and more regular with smaller and smaller topological variation with the concentration increase. And TMAO systems are clearly much more stable compared with urea.

\subsection{Topological modeling for hydrogen-bonding networks}

\begin{figure}
\begin{center}
\begin{tabular}{c}
\includegraphics[width=0.8\textwidth]{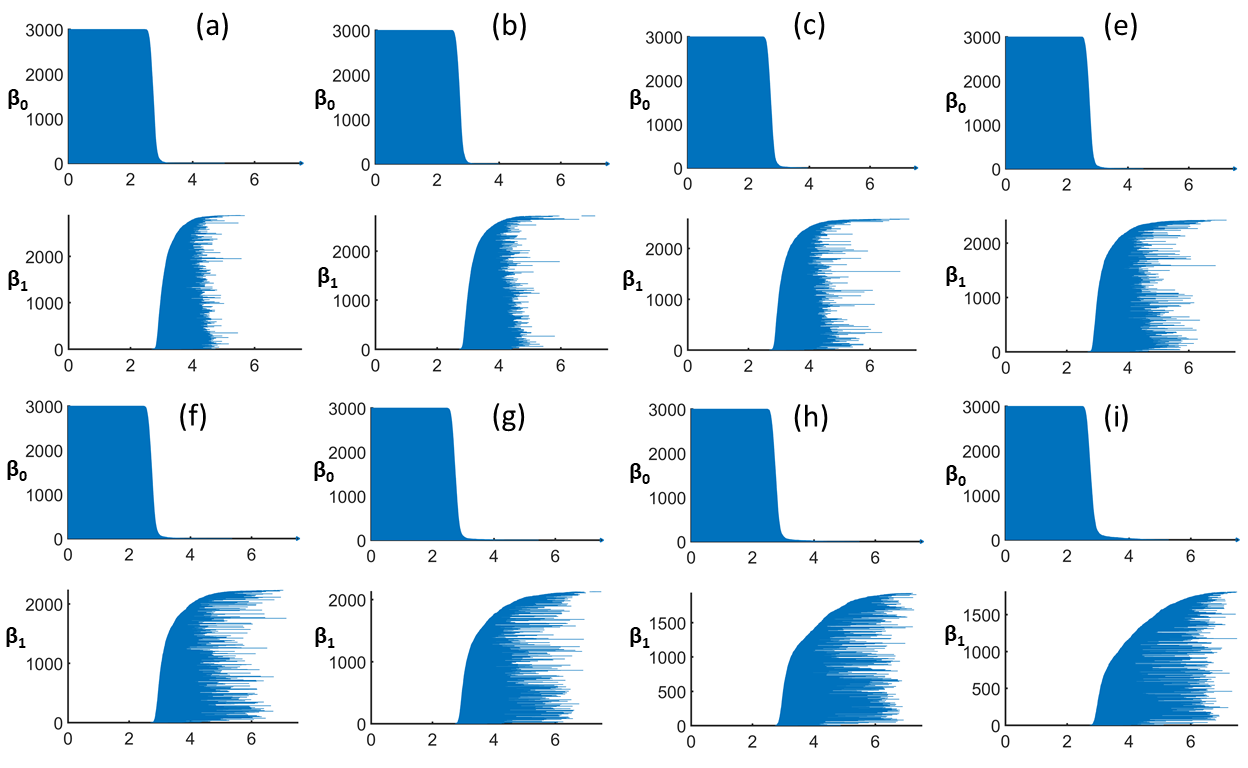}
\end{tabular}
\end{center}
\caption{ The persistent barcodes for hydrogen-bonding networks from TMAO systems. The coarse-grained representation of water as its oxygen atom is considered. The last configuration of the MD simulation is used.  (${\bf a}$) to (${\bf i}$) are persistent barcodes for hydrogen-bonding networks from TMAO systems with ion concentration 1 M to 8 M, respectively. It can be seen that, the total bar numbers of $\beta_0$ barcodes remain as a constant 3000 from (${\bf a}$) to (${\bf i}$), representing the total number of 3000 ${\rm H_2O}$ molecules in all simulations. In the meantime, the $\beta_1$ barcodes decrease from around 2800 to 1700, meaning ring/loop structures become less and less. More importantly, as the concentration increases, more and more longer $\beta_1$ barcodes are generated. To summarize, with the increase of TMAO concentrations, the corresponding hydrogen-bonding networks have less and less circle or loop structures, but the relative size of them increases.
}
\label{fig:tmao_4p_h2o}
\end{figure}

We also consider topological properties of the TMAO and urea hydrogen-bonding networks. Figure \ref{fig:tmao_4p_h2o} illustrates the  persistent barcodes for TMAO hydrogen-bonding networks from eight different concentrations.  The coarse-grained representation of water as its oxygen atom is considered. Similar to the above TMAO and urea molecular aggregation analysis, the last configuration of the MD simulation is used. Figure \ref{fig:tmao_4p_h2o}(${\bf a}$) to (${\bf i}$) are persistent barcodes for hydrogen-bonding networks from TMAO systems with ion concentration 1 M to 8 M, respectively. It can be seen that, the total numbers of $\beta_0$ bars remain as a constant, i.e., 3000, in Figure \ref{fig:tmao_4p_h2o}(${\bf a}$) to (${\bf i}$), representing the 3000 ${\rm H_2O}$ molecules in all simulations. In the meantime, the $\beta_1$ barcodes decrease from around 2800 to 1700, meaning ring/loop structures become less and less. More importantly, as the concentration increases, more and more longer $\beta_1$ barcodes are generated. To summarize, with the increase of TMAO concentrations, the corresponding hydrogen-bonding networks have less and less circle or loop structures, but the average sizes of them increase.

\begin{figure}
\begin{center}
\begin{tabular}{c}
\includegraphics[width=0.8\textwidth]{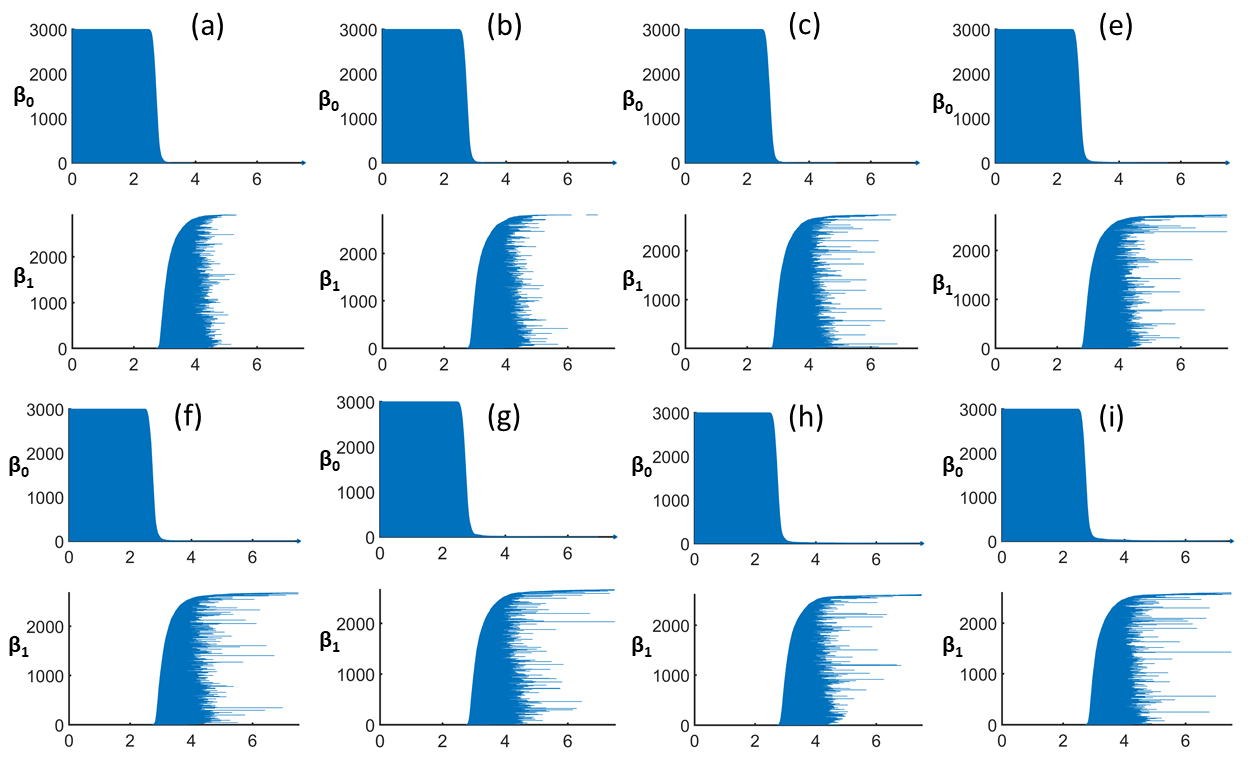}
\end{tabular}
\end{center}
\caption{The persistent barcodes for hydrogen-bonding networks from urea systems. The coarse-grained representation of water as its oxygen atom is considered. The last configuration of the MD simulation is used.  (${\bf a}$) to (${\bf i}$) are persistent barcodes for hydrogen-bonding networks from urea systems with ion concentration 1 M to 8 M, respectively. It can be seen that, the total bar numbers of $\beta_0$ barcodes remain as a constant 3000 from (${\bf a}$) to (${\bf i}$), representing the total number of 3000 ${\rm H_2O}$ molecules in all simulations. Unlike the ones from TMAO systems, the hydrogen-bonding networks from urea systems demonstrate no clear differences in different concentrations.
}
\label{fig:urea_4p_h2o}
\end{figure}

In contrast, the persistent barcodes for hydrogen-bonding networks from urea systems demonstrate different properties. Figure \ref{fig:urea_4p_h2o} demonstrates the barcodes for hydrogen-bonding networks from urea systems with ion concentration 1 M to 8 M. Similar to the TMAO hydrogen-bonding cases, the coarse-grained representation of water as its oxygen atom is considered and only the last configuration of the MD simulation is used.  From Figure \ref{fig:urea_4p_h2o} (${\bf a}$) to (${\bf i}$), it can be seen that, the total numbers of $\beta_0$ bars also remain as the constant 3000. However, unlike TMAO systems, as the concentration increases, the total $\beta_1$ bar numbers for urea systems remain relative stable (at around 2500). At the same time, more and more long $\beta_1$ barcodes appear. It can be noticed that, compared with TMAO, the longer $\beta_1$ bars only account for a very small portion of the $\beta_1$ barcodes. Geometrically, these results mean that the total number of the circle structures in the hydrogen-bonding networks from urea systems remain relatively stable, even though with the concentration increase, a small portion of these circles become much larger.

\begin{figure}
\begin{center}
\begin{tabular}{c}
\includegraphics[width=0.8\textwidth]{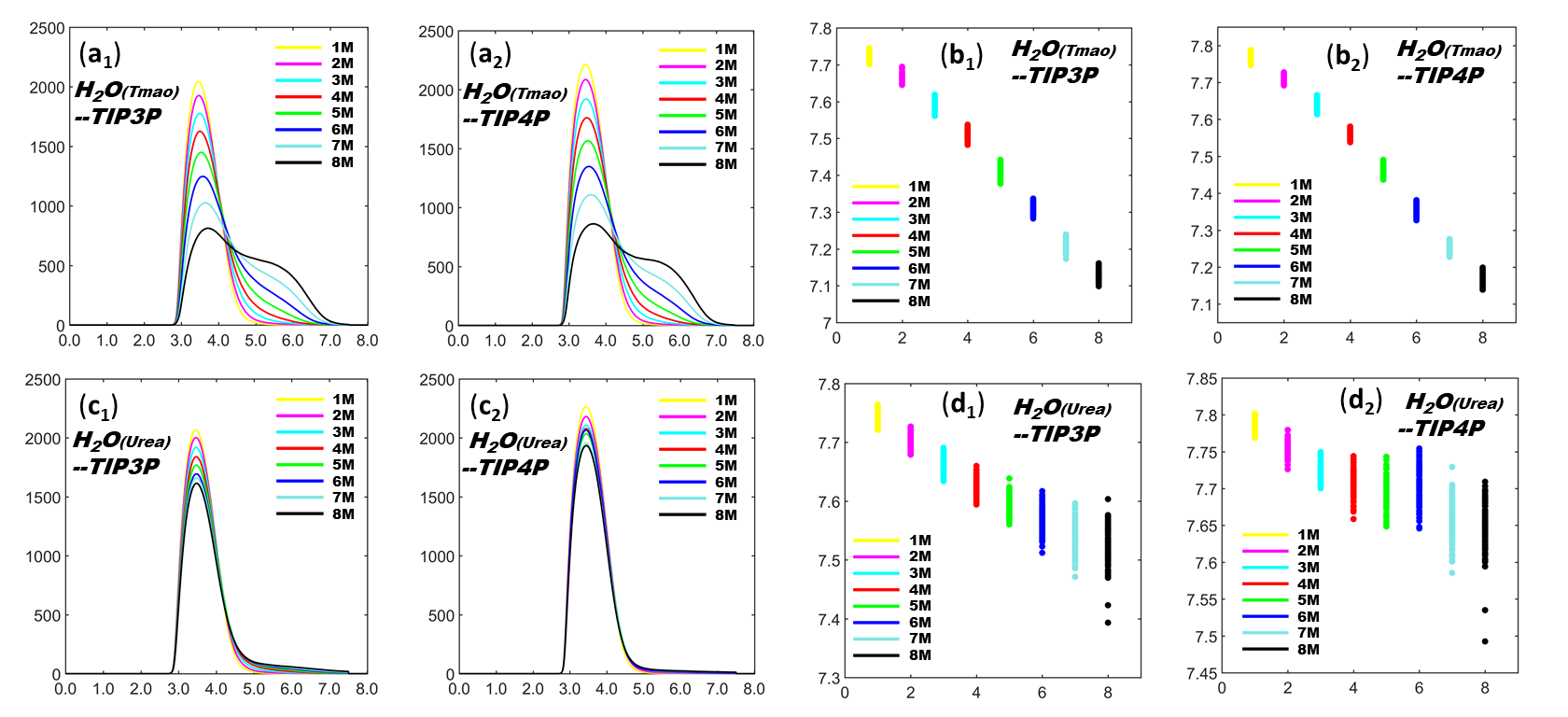}
\end{tabular}
\end{center}
\caption{The comparison of average PBNs and persistent entropies for TMAO and urea hydrogen-bonding networks. Two water models, i.e., TIP3P and TIP4P, are considered. The PBNs are averaged over the total number of atoms and configuration numbers. It can be seen that the PBNs share very similar patterns in both TIP3P and TIP4P models. More importantly, PBNs of the hydrogen-bonding networks from TMAO (${\bf a}$) are dramatically different from those of urea (${\bf c}$). With the concentration increase, the peak value of TMAO hydrogen-bonding PBNs gradually decreases. In the meantime, there is a consistent increase of the PBN values in the range from around 4.2\AA~ to 7.0\AA. In contrast, PBNs from urea hydrogen-bonding networks remains relatively stable with only a slight decrease in their peak values. The persistent entropies for hydrogen-bonding networks from TMAO (${\bf c}$) and urea (${\bf d}$) systems. Totally 101 configurations for each concentration are considered. It can be seen that the PEs share very similar patterns in both TIP3P and TIP4P models. With the concentration increase, average PEs for hydrogen-bonding networks from both TMAO and urea decrease. However, PEs for urea hydrogen-bonding networks decrease much slower from around 7.75 to 7.55. In contrast, PEs for TMAO hydrogen-bonding networks decrease from around 7.75 to 7.10. Moreover, the variance for TMAO systems remains relative stable during the concentration variation, while that of urea systems systematically increases.
}
\label{fig:PBN_h2o}
\end{figure}

After the above case study, we provide a thorough comparison for the hydrogen-bonding networks from TMAO and urea systems. We consider the average PBNs and PEs in the same way as the above molecular aggregation systems. Both TIP3P and TIP4P water models are considered and the results are demonstrated in Figure \ref{fig:PBN_h2o}. Similar to the molecular aggregation, PBNs from both TIP3P and TIP4P models have similar behaviors. More specifically, PBNs for TMAO hydrogen-bonding networks have only slight difference in their peak values. In contrast, PBNs for urea hydrogen-bonding networks show a comparable larger difference in their peaks, even though the general shapes for PBNs are very similar. Further, PBNs for both systems have their peak values at the filtration size around 3.5 \AA. Other than that, PBNs of the hydrogen-bonding networks from TMAO and urea are dramatically different. With the concentration increase, the peak value of TMAO hydrogen-bonding PBNs gradually decreases. In the meantime, there is a consistent rise of the PBN values in the range from around 4.5 \AA~ to 7.0 \AA. In contrast, PBNs from urea hydrogen-bonding networks remain relatively stable with only slight decrease of the their peak values. There is also a very slight increase of PBN values in the range from around 4.5 \AA~ to 7.0 \AA~ with the concentration increase. However, compared with the TMAO, the increase is almost negligible. Geometrically, these results mean that the hydrogen-bonding networks undergo dramatic topological changes for TMAO systems with the concentration increase. More and more large circle structures are formed within the hydrogen-bonding network. In contrast, little damage is done to the hydrogen-bonding network in urea systems with the concentration rise. Only a very small portion of circle structures are generated.

Figure \ref{fig:PBN_h2o} also illustrate the PEs for hydrogen-bonding networks from TMAO and urea systems. For each concentration, totally 101 configurations are considered. It can be seen that the PEs share a roughly similar pattern in both TIP3P and TIP4P models. However different from molecular aggregation systems, the average PEs for both hydrogen-bonding networks consistently decrease with the concentration increase. It should be noticed that TMAO hydrogen-bonding networks have larger PE variations, changing from around 7.7 to 7.1. In contrast, average PEs for urea hydrogen-bonding networks decrease much slower from around 7.8 to 7.5. Moreover, the PEs for the two hydrogen-bonding networks have very different variances. Table \ref{tab:PEs} demonstrates the variances of PEs for both systems. It can be seen that the variance for TMAO and urea all increase with the concentration. However, PE variance for TMAO increase much slower than urea. The ratio between the variance of concentration at 8 M and 1 M for TMAO is around 2 for both TIP3P and TIP4P models. While the ratio for urea is around 15 and 20 for TIP3P and TIP4P models, respectively. In this way, even though TMAO and urea have similar variance at concentration 1 M, they differ greatly at concentration 8 M. The variance for urea PEs is about 7 times larger than that of TMAO at 8 M. These variance results mean that, both TMAO and urea hydrogen-bonding networks become more and more unstable with relatively more topological variation with the concentration increase. But TMAO hydrogen-bonding networks are clearly much more stable compared with urea.

It is worth mentioning that the PBNs for the two types of hydrogen-bonding networks at different concentrations share remarkable similarity with the PBNs from NaCl and KSCN systems\cite{xia2018persistent}. More specifically, hydrogen-bonding networks from NaCl and urea all have only one peak in their PBNs and their peak values decrease with the concentration increase. Hydrogen-bonding networks from KSCN and TMAO all have the two-peak shapes as in Figure \ref{fig:PBN_h2o}. They all have dramatic decrease in the first peak value and gradual increase of the second peak value with the concentration increase. These highly consistent topological features indicate the existence of two topological structures in hydrogen-bonding networks for ``structure making" and ``structure breaking" systems.


\begin{table}[htbp]
  \centering
    \small
	\caption{The comparison of variance of topological entropies for TMAO and urea aggregations and hydrogen-bonding networks. The results are calculated from the MD simulation done with two water models and eight different concentrations.}
    \begin{tabular}{|c|c|c|c|c|c|c|c|c|} \hline
\multirow{2}{*}{C} & \multicolumn{4}{c|}{TIP-3P} & %
    \multicolumn{4}{c|}{TIP-4P} \\
    \cline{2-9}
    & TMAO & urea  & $\rm H_2O(TMAO)$ & $\rm H_2O(urea)$  & TMAO & urea  & $\rm H_2O(TMAO)$ & $\rm H_2O(urea)$\\ \hline
 1 &0.019489   &0.053585   &0.000072   &0.000074   &0.027279   &0.036302   &0.000060   &0.000052  \\ \hline
 2 &0.008980   &0.018623   &0.000081   &0.000103   &0.010533   &0.019381   &0.000064   &0.000080  \\ \hline
 3 &0.004183   &0.011745   &0.000114   &0.000143   &0.004129   &0.011360   &0.000092   &0.000116  \\ \hline
 4 &0.002567   &0.006079   &0.000114   &0.000197   &0.002442   &0.015710   &0.000088   &0.000270  \\ \hline
 5 &0.001812   &0.007377   &0.000176   &0.000206   &0.001550   &0.010768   &0.000107   &0.000460  \\ \hline
 6 &0.000854   &0.007875   &0.000127   &0.000495   &0.000922   &0.019308   &0.000150   &0.000635   \\ \hline
 7 &0.000719   &0.007164   &0.000142   &0.000816   &0.000697   &0.012068   &0.000121   &0.000711   \\ \hline
 8 &0.000365   &0.005135   &0.000151   &0.001094   &0.000399   &0.008485   &0.000160   &0.001012 \\ \hline
    \end{tabular}
  \label{tab:PEs}
\end{table}

\section{Conclusion Remarks}

In this paper, we use the persistent homology to explore the topological properties for osmolyte molecular aggregation and their hydrogen-bonding networks. We have found that TMAO and urea show two extremely different topological behaviors, i.e., extensive networks and local clusters, with the increase of their concentration. More specifically, TMAO tends to form highly consistent large loop or circle structures, while urea will tightly aggregate locally, at same time, form extremely large circles structures in high concentrations. Further, with the concentration increase, TMAO hydrogen-bonding networks vary greatly in their total number of loop structures and large-sized loops consistently increase. In contrast, urea hydrogen-bonding networks remain relatively stable with a slight reduce of the loop numbers and a marginal increase of large-sized loops. We also consider persistent entropy (PE) in the characterization of the topological information of the molecular aggregation and hydrogen-bonding networks. We have found that PEs increase with the concentration for both TMAO and urea, and decreases for their hydrogen-bonding network. More interestingly, the variance of PEs differs significantly for TMAO and urea systems. Finally, we point out that the topological features of the osmolyte molecular aggregations and their hydrogen-bonding networks are highly consistent with those from the ion aggregation systems, which indicates that the topological invariants can characterize very well the intrinsic features of the ``structure making" and ``structure breaking" systems.

\section*{Acknowledgments}
This work was supported in part by Nanyang Technological University Startup Grant M4081842 and Singapore Ministry of Education Academic Research fund Tier 1 RG31/18, Tier 2 MOE2018-T2-1-033.
\vspace{0.6cm}


\end{document}